\documentstyle[12pt,aaspp4]{article}

\newcommand{\msun}{\mbox{$M_{\sun}$}} 
 
\newcommand{\bq}{\begin{equation}} 
\newcommand{\eq}{\end{equation}}

\begin{document} 
\def\refitem{\par\parskip 0pt\noindent\hangindent 20pt} 

\normalsize 

\title{Identification of Type Ia Supernovae at 
Redshift 1.3 and Beyond with the Advanced Camera for Surveys on 
HST\altaffilmark{1}} 
\vspace*{0.3cm} 

{\it  \ \ \ \ \ \ \ \ Accepted for Publication by the Astrophysical Journal Letters}
\vspace*{0.3cm}

Adam G. Riess\altaffilmark{2}, 
Louis-Gregory Strolger\altaffilmark{2}, 
John Tonry\altaffilmark{3}, 
Zlatan Tsvetanov\altaffilmark{2}, 
Stefano Casertano\altaffilmark{2}, 
Henry C. Ferguson\altaffilmark{2}, 
Bahram Mobasher\altaffilmark{2}, 
Peter Challis\altaffilmark{4}, 
Nino Panagia\altaffilmark{2}, 
Alexei V. Filippenko\altaffilmark{5}, 
Weidong Li\altaffilmark{5}, 
Ryan Chornock\altaffilmark{5}, 
Robert P. Kirshner\altaffilmark{4}, 
Bruno Leibundgut\altaffilmark{6}, 
Mark Dickinson\altaffilmark{2}, 
Anton Koekemoer\altaffilmark{2}, 
Norman A. Grogin\altaffilmark{2}, 
and Mauro Giavalisco\altaffilmark{2} 

\altaffiltext{1}{Based on observations with the NASA/ESA {\it Hubble Space 
Telescope}, obtained at the Space Telescope Science Institute, which is 
operated by AURA, Inc., under NASA contract NAS 5-26555} 
\altaffiltext{2}{Space Telescope Science Institute, 3700 San Martin 
Drive, Baltimore, MD 21218} 
\altaffiltext{3}{Institute for Astronomy, University of Hawaii, 
2680 Woodlawn Drive, Honolulu, HI 96822} 
\altaffiltext{4}{Harvard-Smithsonian Center for Astrophysics, 60 Garden St., 
Cambridge, MA 02138} 
\altaffiltext{5}{Department of Astronomy, 601 Campbell Hall, University of 
California, Berkeley, CA  94720-3411} 
\altaffiltext{6}{European Southern Observatory, Karl-Schwarzschild-Strasse 
 2, Garching, D-85748, Germany}

\begin{abstract} 

We present discoveries of SNe Ia at $z > 1$ and the photometric diagnostic
used to discriminate them from other types of SNe detected during the 
GOODS {\it Hubble Space Telescope} Treasury program with the Advanced Camera 
for Surveys (ACS).  Photometric redshift measurements of the hosts combined 
with deep $f606w$, $f775w$, and $f850lp$ imaging discriminates 
hydrogen-rich SNe~II from SNe~I at $z > 1$ by exploiting the ultraviolet (UV) deficit in the energy distributions of SNe~Ia.  This sorting allows rapid follow-up of space-based discoveries.  Subsequent 
spectroscopy of 11 GOODS SNe~Ia obtained from the ground and with the grism on ACS 
confirmed the reliability of our photometric screening.  We 
present the highest-redshift spectrum of any supernova published to date, SN~Ia 2002fw at $z = 1.3$ 
observed near maximum brightness with the ACS grism.  The lack of UV flux for true SNe~Ia provides an effective tool for our ongoing efforts to build a sample of 
SNe~Ia at $1.2 < z < 1.8$ which will extend the useful range of the 
magnitude-redshift relation of SNe~Ia.

\end{abstract} 
subject headings:  supernovae: general --- cosmology: observations 

\section{Introduction} 

The unexpected faintness of Type Ia supernovae (SNe~Ia) at $z \approx 0.5$ provides direct evidence that 
the expansion of the Universe is accelerating, propelled by ``dark energy'' 
(Riess et al.  1998; Perlmutter et al. 1999).  Augmented samples of SNe~Ia 
extended to $z \approx 1$ (Tonry et al. 2003; Suntzeff et al. 2003; Leibundgut 
et al. 2003; Blakeslee et al. 2003), as well as more detailed studies of SNe Ia 
(see Leibundgut 2001 and Riess 2000 for reviews), have strengthened the case 
for the existence of dark energy based on supernovae.  The new cosmic concordance model 
contains both dark matter and dark energy (i.e., $\Omega_M \approx 0.3$ and 
$\Omega_\Lambda\approx0.7$) and is sharply delineated by the convergence 
of three independent probes of the $\Omega_M-\Omega_\Lambda$ plane: 
the cosmic 
microwave background (CMB), SNe~Ia at $z < 1$, and large-scale structure (LSS). However, when the number of constraints nearly equals the number of parameters little power remains to test the validity 
of the paradigm.  Efforts to use the same measurements
to fit an additional parameter,
 such the equation of state for dark energy, are vulnerable to systematic errors without redundant measurements. 

A simple prediction of the new cosmological model is the weakening of dark 
energy relative to dark matter in proportion to $(1+z)^3$, leading to 
an earlier epoch dominated by dark matter, with decelerating expansion.  
If we fail to see deceleration, the evidence for dark energy is in doubt.   The tools for this test 
are SNe~Ia at $z > 1$, whose leverage to measure past deceleration 
grows with redshift.  SNe~Ia at $z > 1$ also create the opportunity to {\it amplify} the effects of any astrophysical 
contaminants such as chemistry, age, or strange dust that might contaminate the dark energy signal from SNe~Ia. 

To date, the reach of discovery from the ground has been limited to $z < 1.2$ 
(Aldering et al. 1998; Coil et al. 2000; Tonry et al. 1999, 2003).  Pursuit of SNe~Ia to higher redshifts is inhibited by the 
bright night sky redward of $\sim$8000~\AA.  Searches at bluer 
wavelengths are starved of photons by the intrinsic deficiency of SNe~Ia in the UV.  Even at $1.0 < z < 1.2$, detections of SNe~Ia from the ground 
are often not reliable.  Success requires unusually favorable conditions and daringly low significance thresholds for detections, but attempts can result in false 
discoveries, inconclusive spectra, or incomplete photometric records 
(e.g., SN 1999fo, SN 1999fu and SN 1999fv; Tonry et al. 2003).  Spectroscopic screening of 
low-confidence SN~Ia candidates at $z > 1$ is always time-consuming and often impractical and 
unsuccessful. 

From space these challenges are largely mitigated.  Gilliland \& Phillips 
(1998) reported the detection of SN 1997ff by WFPC2 on the Hubble Space 
Telescope {\it (HST)}.  Using serendipitous follow-up of the field by NICMOS, 
Riess et al. (2001) showed that this SN was almost certainly of type Ia at $z 
\approx 1.7$ and its apparent magnitude consistent with deceleration.  However, a single, sparsely 
observed SN is insufficient to guard against a spurious conclusion caused by 
lensing (Ben\'\i tez et al. 2002) or an unexpected SN type. 

An unprecedented opportunity to collect a significant sample of SNe~Ia at $1 < 
z < 2$ occurred in the eleventh year of {\it HST} owing to three factors: the 
installation of a wider and faster ACS (Ford et al. 2003), the refurbishment of 
NICMOS, and the periodic (45 day interval) multi-color imaging of two $10' \times 
15'$ fields with ACS by the GOODS Treasury Program (Giavalisco et al. 2003). 
Each of the 8 repeated epochs of the GOODS imaging was expected to yield 1--2 
SNe~Ia at $z > 1.2$ as well as 3-5 other SNe.  The ability to discriminate 
the SNe~Ia at $z > 1.2$ from other transients with high confidence is essential 
to avoid following the wrong objects, or squandering spectroscopic time and to permit prompt 
{\it HST} monitoring via a pre-allocated program .  The combined GOODS and SN 
``piggyback'' program (GO 9352; Riess, PI) constitutes the first program designed to discover and follow SNe from space. 

We have used observed colors and photometric redshifts to identify likely 
SNe~Ia at $z > 1$ via their intrinsic UV deficit.  Subsequent measurements
 using the grism 
on ACS have provided the spectroscopic confirmation at redshifts
unreachable from the ground including our first example and
 the highest redshift 
supernova yet measured, in this case a SN~Ia at $z = 1.3$.  Here we report on a novel 
photometric method used to identify candidate SNe~Ia at $z > 1$ and demonstrate its 
spectroscopic confirmation.  Elsewhere we report details of the 
search for transients and their rates (Strolger et al. 2003; Dahlen et 
al. 2003c), as well as the constraints on cosmology and the nature of distant 
SNe~Ia (Riess et al. 2003).

\section{Identifying SNe~Ia at $z > 1$; the UV Deficit} 

    The modern classification scheme of SNe includes at least 7 primary 
subtypes (SN~IIL, IIP, IIb, IIn, Ia, Ib, and Ic) identified via the presence or 
absence of key features in their optical spectra (see Filippenko 1997 for a 
review).  The major dichotomy of classification, separates SNe with and without hydrogen (SNe~II 
and I, respectively) in their spectra.  Unfortunately, this historical 
convention fails to reflect current understanding which recognizes a different, 
more physical bifurcation, that of explosion mechanism.  SNe~Ia are believed to 
arise from the thermonuclear disruption of a white dwarf 
and all other subtypes via core collapse in a massive ($\gtrsim 10 \msun$) star with varying degrees of mass loss
(Wheeler \& Harkness 1990; Filippenko 1997). 

Explosive thermonuclear burning of the degenerate SN~Ia progenitor produces 
0.3--1 $\msun$ of iron-group metals. Owing to 
its high velocities and great abundance, Fe~II resonantly scatters much of the 
thermal continuum flux in the UV ($\lambda < 3300$~\AA), imprinting a 
characteristic ``UV deficit'' in the spectral energy distributions of SNe~Ia. 
In contrast, the most common form of core-collapse SNe, those with 
hydrogen-rich, electron-scattering atmospheres, are strong emitters in the UV. 
As shown in Figure 1, the UV deficit of SNe~Ia provides a useful, photometric 
method to discriminate between the two most frequently observed subtypes of SN, SNe~Ia and SNe~II. 
Rest-frame optical ($\lambda > 3300$~\AA) color discrimination of SNe has been 
shown by Poznanski et al. (2002) to be of use for this purpose, though photometric differences 
are more subtle at these wavelengths than in the UV. 

Although SNe~Ia may be routinely discovered and identified via photometric 
means, misidentification is possible.  Core-collapse SNe with negligible 
hydrogen (SNe~Ib and Ic) also have weak UV flux.  However, such SNe are less 
common in the field than hydrogen-rich core-collapse SNe by a factor of $\sim 
3$ (Cappellaro et al. 1997; Cappellaro, Evans, \& Turatto 1999; W. Li and 
A. V. Filippenko, 2003, private communication), and most are 
fainter than SNe~Ia by 1--3 mag, effectively reducing their discovery rate in a 
magnitude-limited survey (e.g., Li et al. 2001).  More troublesome is a subset 
of SNe~Ic which attain optical luminosities comparable to those of SNe~Ia and would be
 the chief contaminant to SNe~Ia 
identified in magnitude-limited surveys in the optical.  However, past surveys 
of this nature demonstrate that the degree of contamination is small. 
Ultra-luminous SNe~Ic constitute only a few percent of SNe found by the 
Cal\'{a}n-Tololo Survey (Hamuy et al. 1996; Clocchiatti et al. 2000), the Abell 
Cluster Search (Germany et al. 2003), the Lick Observatory 
Supernova Search (Filippenko et al. 2001; W. Li and A. V. Filippenko, 2003, 
private communication), and the Nearby SN Factory Search (P. Nugent, 2003, 
private communication), or in compilations of past surveys (Richardson et 
al. 2002).  Heavily reddened core-collapse SNe could mimic the colors of a 
SN~Ia in the UV, but the total absorption would severely reduce 
their detection rate at $z > 1$.  Without a significant evolution 
in the relative rates of SN types or their photometric properties, the expected 
frequency of misidentification of candidate SN~Ia targeted by this diagnostic would be less 
1 in 10, although the rate may rise at higher redshifts.
   Because cosmological measurements from SNe Ia are statistical in nature, modeling of the frequency of contamination can be used to correct a sample but the best defense against sample contamination is spectroscopic screening.

The GOODS Survey provides the means to use this 
photometric discriminant for SNe~Ia at $z > 1$.  SNe in the GOODS survey were 
discovered by image subtraction in 2000~s exposures in ACS $f850lp$, a passband 
that includes the rest-frame optical ($UBVRI$) up to $z \approx 1.5$ and 
allows for the detection of SNe~Ia up to $z \approx 1.8$.  Simultaneous imaging 
in $f775W$ and $f606w$ provides a useful measurement of the UV flux at $z > 1$. 
As shown in Figure 2, the expected $f775w - f850lp$ and $f606w - f850lp$ colors 
of SNe~Ia and SNe~II diverge at $z > 1.0$ and $z > 0.7$, respectively, with 
apparent differences reaching 1--3 mag.  SNe~Ia at $z > 1.3$ are expected to be 
so faint in $f606w$, that such SNe 
should appear as ``drop-outs.''  The survey limit of $f850lp \approx 25.8$ mag 
(see Strolger et al. 2003 for more precise values) restricts the discovery of 
SNe at $z > 1.2$ to $M_B < -17.5$ mag and at $z > 1.5$ to $M_B < -18.5$ mag. 

Within 6 to 18 hours after a repeat visit of the Chandra Deep Field South or Hubble Deep Field North, each GOODS epoch 
was differenced with the preceding epoch 
to identify SNe.  Follow-up spectroscopy of most candidates 
with $f850lp < 24$ mag was attempted from the ground with the Keck and Magellan 
telescopes during the next new moon and are reported elsewhere (Riess et 
al. 2003).  Candidates whose host photometric redshift indicated a high likelihood of 
$z > 1.2$ and 
whose colors indicated a good match to the UV deficit of an SN~Ia at the host 
redshift were targeted for follow-up with {\it HST} using ACS direct imaging, 
NICMOS, and in some cases ACS grism spectroscopy.  All likely SNe were reported 
in the IAU Circulars within 72 hours of discovery, 43 to date (Giavalisco et al. 2002; 
Riess et al.  2002, 2003; Strolger et al. 2002; Casertano et al. 2003; Ferguson et 
al. 2003; Dahlen et al. 2003a).  A complete listing of the GOODS SNe and their 
classifications can be found in Strolger et al. (2003).  Here we focus on the 
initial identification via the apparent UV deficit 
of 15 candidate SNe~Ia in three redshift intervals: $z < 1.0$, $1 
< z <1.5$, and $z > 1.5$. 

\section{High, Higher, Highest} 

For four SNe~Ia with $0.2 < z < 1$, the apparent magnitude of $21 < f850lp < 
23.5$ mag made it possible to get a spectroscopic confirmation in a few hours 
on a 6~m to 10~m telescope (three others were confirmed
 in ACS grism images).  Each of these spectra provide conclusive 
identification as a SN~Ia, as well as the indicated redshift, and are presented 
elsewhere (Strolger et al. 2003; Riess et al. 2003).  In all cases the measured 
photometric redshifts and colors were consistent with the spectroscopic 
identification as shown in Figure 2.  However, because the effective rest-frame 
wavelength of the $f606w$ bandpass remains redward of 3300~\AA\ at $z < 1$, the 
observed colors do not sample the UV deficit and were therefore insufficient to 
distinguish the transients as probable SNe~Ia from colors alone. 

For five transients, the observed colors and the photometric redshifts of the 
hosts were consistent with being SNe~Ia at $1 < z < 1.5$ and inconsistent with 
being SNe~II.  In this redshift interval the observed colors were highly 
indicative of SN type, allowing us to select these as likely SNe~Ia for target 
of opportunity (ToO) monitoring with {\it HST} in advance of spectral confirmation and without delays or 
gaps in the photometric record.  Four of the five were spectroscopically confirmed as 
SNe~Ia at a redshift consistent with the initial photometric estimate as shown 
in Figure 2 (Riess et al. 2003). 

The deep imaging obtained by GOODS before the discovery of SN 2002fw indicated 
$1.1 < z < 1.5$ with 95\% confidence for its host galaxy.  The measured colors 
of SN 2002fw at discovery of $f775w - f850lp = 0.80 \pm 0.05$ mag and $f606w - 
f850lp = 3.0 \pm 0.1$ mag were consistent with a SN~Ia at the photometric 
redshift (but not a SN~II), precipitating a prompt {\it HST} 
monitoring (ToO).  In Figure 3 we present the {\it HST} spectroscopy of SN 
2002fw, a SN~Ia at $z = 1.3$ and the first high-redshift SN~Ia harvested from 
the GOODS program (see \S 3).  The spectrum was obtained in 15~ks of exposure with the ACS 
WFC and the grism filter, beginning 10 days after the discovery image.  Similar ACS grism 
data 
was obtained for other SNe Ia in this redshift range and will be shown elsewhere (Riess et al. 2003).

Two SNe were discovered whose photometric redshifts and colors were indicative 
of SNe~Ia at $z > 1.5$.  In both cases the $f775w - f850lp$ color was 
between 2.0 and 2.5 mag, consistent with the UV deficit of SNe~Ia and strongly 
inconsistent with the bluer colors of SNe~II.  The absence of either SN in the 
$f606w$ images to the detection limit of point sources (with known position) 
limits $f606w - f850lp > 3.2$ mag for each, consistent with SNe Ia and 
inconsistent with SNe II.  Elsewhere we present the follow-up observations of 
these SNe and provide further constraints on their type and distance (Riess et 
al. 2003). 

\section{Discussion} 

    The UV deficit is a useful tool to identify likely SNe~Ia for 
rapid follow-up observations.  For the GOODS survey careful consideration was given 
to select passbands which could use this diagnostic to discriminate SNe~I 
from SNe~II at $z > 1$.  Monte Carlo simulations of the selection of SNe~Ia at 
$z > 1$ indicate that the frequency of 
``false positives'' (i.e., non-SNe~Ia or SNe~Ia at $z < 1$) would be $<$10\% 
for expected transients.  Further, in each actual case, we can calculate the likelihood of a 
misidentification taking into account the photometric 
measurements, their uncertainties, and the quality of the fit.  The expectedly 
high level of success in identifying SNe~Ia at $z > 1$ allowed us to select 
suitable candidates and commit to rapid follow-up with {\it HST} in advance of 
any additional confirmation.  The subsequent spectroscopy confirmed 
statistically and on an individual basis the good likelihood of selecting 
SNe~Ia via their UV deficit.   

A few SNe were discovered at $z>0.7$ whose blue color
could be used to reject a classification of SN Ia, and 
matches well with classification as a young 
SN II (e.g., $f606w$-$f850lp$=0.0 to 1.0 mag and $z=0.9$; see figure 2).  
However
the signal-to-noise ratio required to confirm spectroscopically
the photometric evidence for young candidate SNe II at these redshifts far exceeds that of 
SNe Ia due both to weakness of features and the candidate's relative faintness and we did not get enough data to confirm this classification.

Due to its high redshift, the well-measured part of the spectrum of SN 2002fw 
extends from 2500~\AA\ to 4200~\AA\ (see Figure 3).  For comparison we have 
superimposed the spectrum of the prototypical SN~Ia 1981B at maximum extended 
to the UV by {\it IUE} (Branch et al. 1985).  Overall the agreement between the 
two is excellent both in color and for specific features.  Only minor 
differences are apparent, consistent with the intrinsic variation of SN~Ia 
abundances and explosion velocities.  Notable and pronounced absorption 
features in the spectrum of SN 2002fw include the \ion{Ca}{2} H\&K feature at 
rest-frame 3700~\AA\ and the blueshifted \ion{Si}{2} $\lambda$4130.  In the UV 
region of the spectrum, a rise blueward of 3200~\AA\ is well discerned due to 
blending of high-velocity \ion{Fe}{2} absorption.  Our line identifications are 
from Kirshner et al. (1993), and Mazzali et al. (1993). 
The defining feature of SNe~Ia, \ion{Si}{2} $\lambda$6355 (typically blueshifted 
to 6150~\AA) cannot be discerned due to the high redshift of the SN.  However, 
as discussed by Coil et al. (2000) for SN 2002fv ($z = 1.2$), the absorption 
feature near 4000~\AA\ due to \ion{Si}{2} $\lambda$4130 is sufficient to 
establish SN 2002fw as a SN~Ia (though not as useful at later times as \ion{Si}{2} $\lambda$6355), and not a SN~Ib/Ic supernova. 

Light curves are a critical component of measuring distances to SNe~Ia.  They 
constrain the age of SN~Ia observations, an individual calibration of each supernova's luminosity,
and the extinction of each SN Ia (via the color excess), 
a prerequisite for calibrating their distances. 
 For $z > 1$ objects, this requires observations in the near-IR with NICMOS. Elsewhere 
(Riess et al. 2003), we present the distances measured for the GOODS SNe Ia 
based on their light curves and colors, as well as the implied cosmological 
constraints. 

\bigskip 
\medskip 
We thank resident and visiting supernova searchers including Tomas Dahlen, Ray Lucas,
Norman Grogin, Peter Garnavich, Ann Hornshmeier, Lexi Moustakas, Marin Richardson,
Ned Taylor, Claudia Kretchmer, Rafal Idzi, 
Kyoungsoo Lee, Richard Hook, Vicki Laidler, Carl Biagetti, 
Duilia de Mello, Swara Ravindranath, and Brian Schmidt. We also thank the steady and heroic
efforts by OPUS, Dorothy Fraquelli, William Januszewski, Mark Calvin, Mark Kochte,
Tracy Ellis, and Bill Workman.
                                                                              
Financial support for this work was provided by NASA through programs 
GO-9352 and GO-9583 from the Space Telescope Science Institute, which is operated by AURA, Inc., under NASA 
contract NAS 5-26555.  Some of the results presented herein were
 made possible by the generous financial support of the 
W. M. Keck Foundation. 
  
\refitem Aldering, G., et al. 1998, IAU Circ. 7046 
\refitem Baron, E., et al. 2000, ApJ, 545, 444
\refitem Ben\'\i tez, N., Riess, A., Nugent, P., Dickinson, M., 
 Chornock, R., \& Filippenko, A. V. 2002, ApJ, 577, L1 
\refitem Blakeslee, et al. 2003, ApJ, in press (astro-ph/0302402) 
\refitem Branch, D., Doggett, J. B., Nomoto, K., \& Thielemann, F.-K. 
 1985, ApJ, 294, 619 
\refitem Cappellaro, E., Evans, R., \& Turatto, M. 1999, A\&A, 351, 459 
\refitem Cappellaro, E., Turatto, M., Tsvetkov, D. Yu., Bartunov, O. S., 
 Pollas, C., Evans, R., \& Hamuy, M. 1997, A\&A, 322, 431 
\refitem Casertano, S., et al. 2003, IAU Circ. 8052 
\refitem Clocchiatti, A., et al. 2000, ApJ, 529, 661 
\refitem Coil, A. L., et al. 2000, ApJ, 544, L111 
\refitem Dahlen, T., et al. 2003a, IAU Circ. 8081 
\refitem Dahlen, T., et al. 2003c, in preparation 
\refitem Ferguson, H. C., et al. 2003, IAU Circ. 8069 
\refitem Filippenko, A. V. 1997, ARAA, 35, 309 
\refitem Filippenko, A. V., Li, W. D., Treffers, R. R., \& Modjaz, 
  M. 2001, in Small-Telescope Astronomy on Global Scales, ed. 
  W. P. Chen, C. Lemme, \& B. Paczy\'{n}ski (San Francisco: ASP), 121 
\refitem Ford, H., et al. 2003 (ACS paper) 
\refitem Germany, L. M., et al. 2003, A\&Ap, in press 
\refitem Giavalisco, M., et al. 2002, IAU Circ. 7981 
\refitem Giavalisco, M., et al. 2003 (GOODS paper) 
\refitem Gilliland, R. L., \& Phillips, M. M. 1998, IAU Circ. 6810 
\refitem Hamuy, M., Phillips, M. M., Maza, J., Suntzeff, N. B., 
 Schommer, R. A., \& Aviles, R. 1996, AJ, 112, 2391 
\refitem Kirshner, R. P., et al. 1993, ApJ, 415, 589 
\refitem Leibundgut, B. 2001, ARAA, 39, 67 
\refitem Leibundgut, B., et al. 2003, in preparation 
\refitem Lentz et al., 2001, ApJ, 547, 406
\refitem Li, W., Filippenko, A. V., Treffers, R. R., Riess, 
 A. G., Hu, J., \& Qiu, Y. 2001, ApJ, 546, 734 
\refitem Mazzali, P. A., Lucy, L. B., Danziger, I. J., Gouiffes, C., 
 Cappellaro, E., \& Turatto, M. 1993, A\&Ap, 269, 423 
\refitem Pain, R., et al. 1996, ApJ, 473, 356 
\refitem Perlmutter, S., et al. 1999, ApJ, 517, 565 
\refitem Poznanski, D., Gal-Yam, A., Maoz, D., Filippenko, A. V., 
 Leonard, D. C., \& Matheson, T. 2002, PASP, 114, 833 
\refitem Richardson, M., et al. 2002, xxxxx 
\refitem Riess, A. G. 2000, PASP, 112, 1284 
\refitem Riess, A. G., et al. 1998, AJ, 116, 1009 
\refitem Riess, A. G., et al. 2001, ApJ, 560, 49 
\refitem Riess, A. G., et al. 2002, IAU Circ. 8012 
\refitem Riess, A. G., et al. 2003, in preparation 
\refitem Schmidt, B. P., et al. 1998, ApJ, 507, 46 
\refitem Strolger, L.-G., et al. 2002, IAU Circ. 8038 
\refitem Strolger, L.-G., et al. 2003, in preparation 
\refitem Suntzeff, N., et al. 2003, in preparation 
\refitem Tonry, J., et al. 1999, IAU Circ. 7312 
\refitem Tonry, J. L., et al. 2003, ApJ, submitted 
\refitem Wheeler, J. C., \& Harkness, R. P. 1990, Rep. Mod. Phys. 53, 1467

\vfill \eject 

Figure Captions

Figure 1:Upper Panel: High-resolution spectra of three representative young SNe; SN Ia 1992A (Kirshner et al. 1993)
and a blue SN II (SN 1998S; Lentz et al. 2001) and a redder SN II (SN 1999em; Baron et al. 2000) from 1200 \AA\ to 7000 \AA\, normalized in the optical.  Lower Panel: low-resolution sampling of the same spectra showing
the color differences in the UV.  Fe~II resonantly scatters much of the 
thermal continuum flux of SNe Ia in the UV ($\lambda < 3300$~\AA), imprinting a 
characteristic ``UV deficit'' not seen in the hydrogen-rich photospheres
of common SNe II.

Figure 2: Finding and classifying SNe Ia with ACS via the UV deficit.
The expected $i-z$ (i.e., $F775W-F850lp$) and $v-z$ (i.e., $F606W-F850lp$) colors of SNe~Ia and SNe~II near maximum
brightness (calculated from the SEDs in figure 1) compared to the observed colors of candidate SNe Ia found in GOODs data.
SNe~Ia are readily distinguishable from SNe~II by
their red colors. Reddened SNe~II would be too faint for this magnitude-limited sample.
  SNe shown at $z<1.5$ were independently confirmed to be SNe Ia.

Figure 3: Spectrum of SN Ia 2002fw at $z=1.3$ as observed in 15ks through the grism filter
with ACS on {\it HST} compared to SN Ia 1981B at maximum light.  Identifiable absorption 
features in the spectrum of SN 2002fw include the \ion{Ca}{2} H\&K absorption blend at 
rest-frame 3700~\AA\ and the blueshifted \ion{Si}{2} $\lambda$4130 absorption.

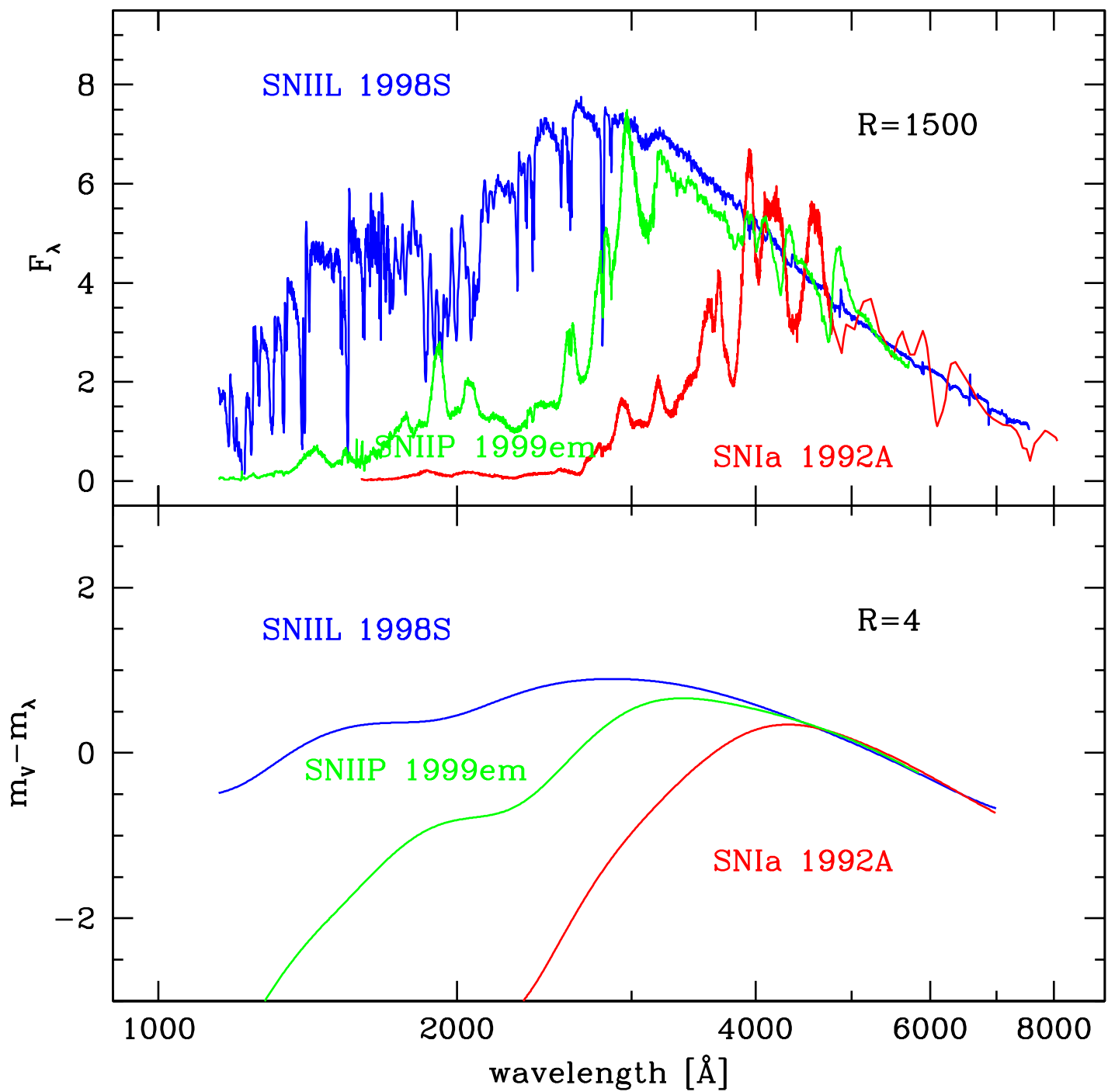
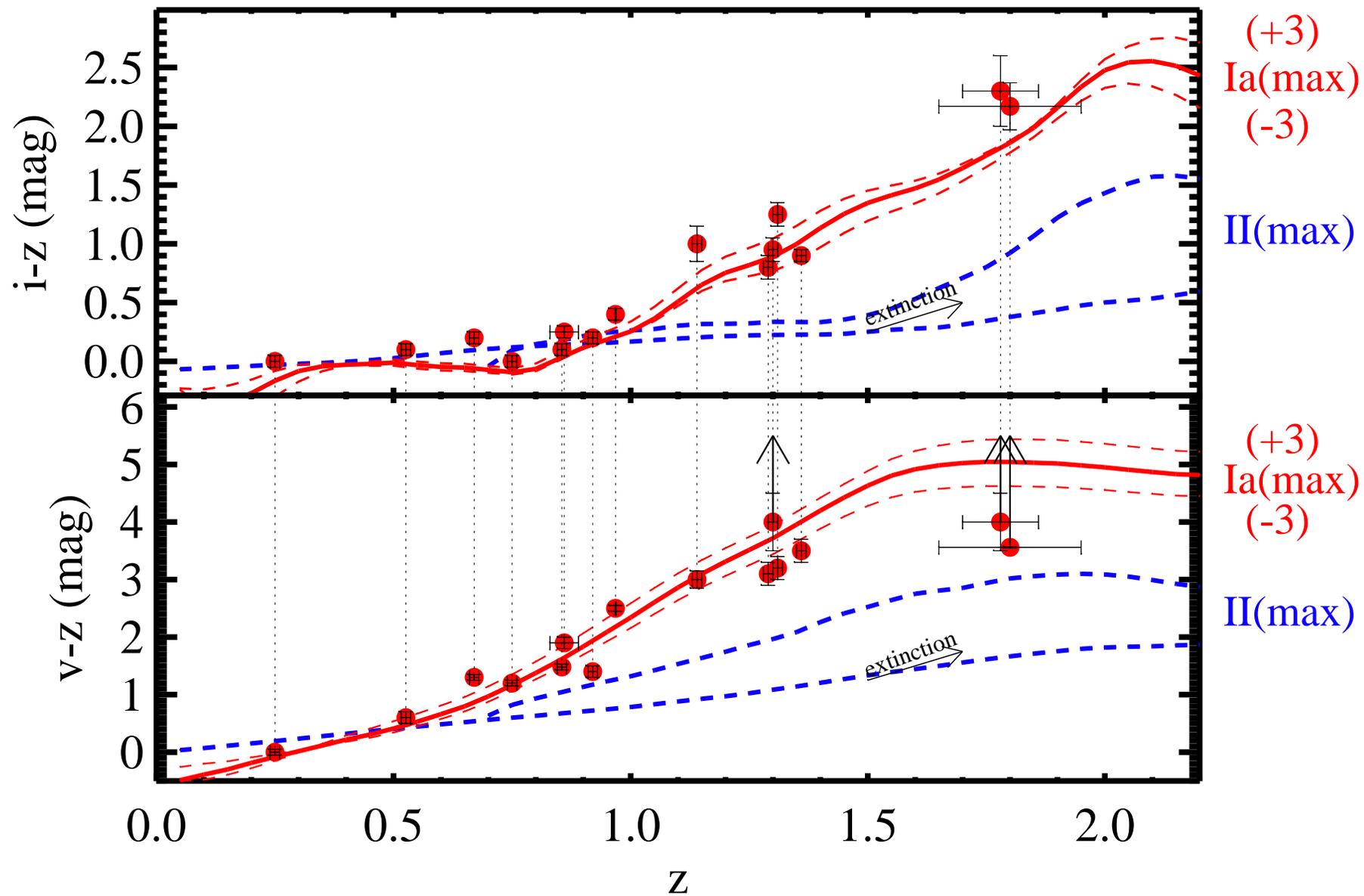
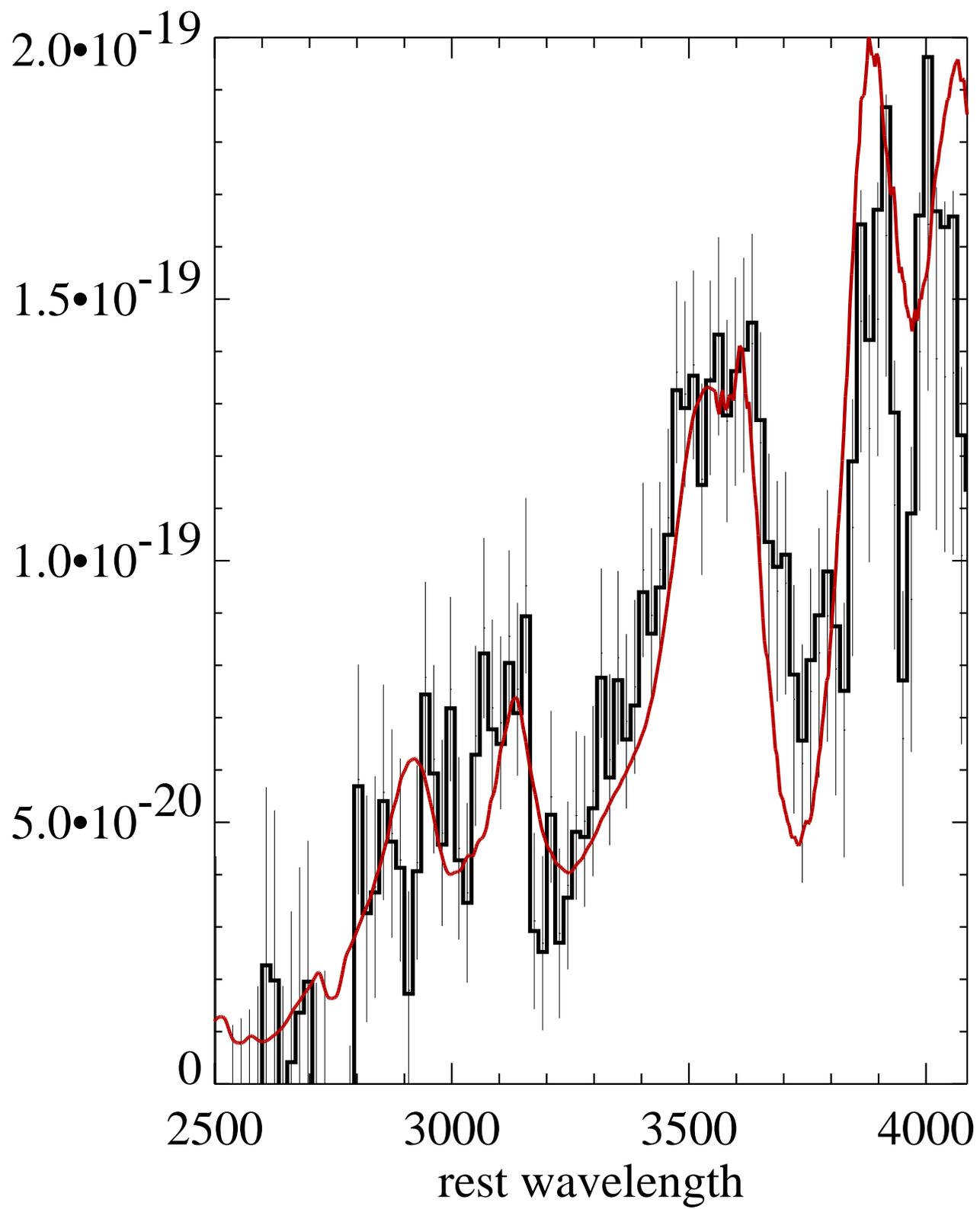
\end{document}